\documentclass[conference]{IEEEtran}
\usepackage{epsfig}
\usepackage{times}
\usepackage{float}
\usepackage{afterpage}
\usepackage{amsmath}
\usepackage{amstext}
\usepackage{amssymb,bm}
\usepackage{latexsym}
\usepackage{color}
\usepackage{graphicx}
\usepackage{amsmath}
\usepackage{amsthm}
\usepackage{graphicx}
\usepackage[center]{caption}
\usepackage{pstricks}
\usepackage{caption}
\usepackage{subcaption}
\usepackage{booktabs}
\usepackage{multicol}
\usepackage{lipsum}% dummy text
\usepackage[T1]{fontenc}
\usepackage{epstopdf}

\allowdisplaybreaks

\newtheorem{thm}{Theorem}%[section]
\newtheorem{cor}{Corollary}

\theoremstyle{definition}

\providecommand{\definitionname}{Definition}

\newfloat{algorithm}{tbp}{loa}
\providecommand{\algorithmname}{Algorithm}
\floatname{algorithm}{\protect\algorithmname}

\long\def\comment#1{}

\newcommand{\dv}{{\mathbf d}}

\newcommand{\Zm}{{\mathbf Z}}

\newcommand{\Hc}{{\mathcal H}}

\newcommand{\Jc}{{\mathcal J}}

\newcommand{\Rc}{{\mathcal R}}

\newcommand{\Uc}{{\mathcal U}}
\newcommand{\Wc}{{\mathcal W}}

\newcommand{\Yc}{{\mathcal Y}}
\newcommand{\Zc}{{\mathcal Z}}

\newcommand{\ksf}{{\mathsf k}}

\newcommand{\nsf}{{\mathsf n}}

\newcommand{\rsf}{{\mathsf r}}

\newcommand{\Bsf}{{\mathsf B}}

\newcommand{\Hsf}{{\mathsf H}}

\newcommand{\Ksf}{{\mathsf K}}

\newcommand{\Msf}{{\mathsf M}}
\newcommand{\Nsf}{{\mathsf N}}

\newcommand{\Rsf}{{\mathsf R}}

\newcommand{\be}{\begin{equation}}
\newcommand{\ee}{\end{equation}}
\newcommand{\bea}{\begin{eqnarray}}
\newcommand{\eea}{\end{eqnarray}}

\begin{document}

\title{A Novel Asymmetric Coded Placement in Combination Networks with end-user Caches}

\author{
\IEEEauthorblockN{%
Kai~Wan\IEEEauthorrefmark{1}, %
Daniela~Tuninetti\IEEEauthorrefmark{3}, %
Mingyue~Ji\IEEEauthorrefmark{2}, %
Pablo~Piantanida\IEEEauthorrefmark{1}, %
}
\IEEEauthorblockA{\IEEEauthorrefmark{1}L2S CentraleSupélec-CNRS-Université Paris-Sud, %Gif-sur-Yvette  91190,
France, \{kai.wan, pablo.piantanida\}@l2s.centralesupelec.fr}%
\IEEEauthorblockA{\IEEEauthorrefmark{3}University of Illinois at Chicago, Chicago, %IL 60607, 
USA, danielat@uic.edu}%
\IEEEauthorblockA{\IEEEauthorrefmark{2}University of Utah, Salt Lake City, %UT 84112, 
USA,  mingyue.ji@utah.edu}%
}

\maketitle
%\IEEEpeerreviewmaketitle{}

\begin{abstract}
The tradeoff between the user's memory size and the worst-case download time in the {\it $(\Hsf,\rsf,\Msf,\Nsf)$ combination network} is studied, where a central server communicates with $\Ksf$ users through $\Hsf$ immediate relays, and each user has local cache of size $\Msf$ files and is connected to a different subset of $\rsf$ relays. 
The main contribution of this paper is the design of a coded caching scheme with {\it asymmetric coded placement} by  
leveraging coordination among the relays, which was not exploited in past work.
Mathematical analysis and numerical results show that the proposed schemes outperform existing schemes. 
\end{abstract}

\section{Introduction}
\label{sec:intro}
%The network serving period can be divided into peak-traffic and off-peak hours. 
Caching is an effective way to smooth out network traffic by storing some contents in users' memories during off-peak times to reduce the required number of transmissions during peak-traffic times. A caching scheme includes two phases. 
{\it In the placement phase}, each user stores parts of content in his cache without knowledge of later demands. 
If each user directly stores some bits of the files, the placement is said to be uncoded. %; otherwise it is coded.
{\it In the delivery phase}, each user
requests one file. According to users' demands and cache contents, the server aims to transmit the smallest number packets so as to satisfy the users' demands, regardless of the demands.

Caching was originally studied by Maddah-Ali and Niesen (MAN) in~\cite{dvbt2fundamental} for the {\it shared-link} network, which comprises a server with $\Nsf$ files, $\Ksf$ users with a cache of size $\Msf$ files, and an error-free broadcast link.
An additional multiplicative {\it coded caching gain} was shown to be attainable by coded caching compared to conventional uncoded caching schemes. For each $\Msf=\Nsf t/\Ksf$, where $t$ is an integer from $0$ to $\Nsf$,  each file is split into $\binom{\Ksf}{t}$ non-overlapping equal-size subfiles that are  strategically placed into the user caches. During the deliver phase, {\it coded multicast messages} are sent through the shared-link so that a single transmission simultaneously serves $t+1$ users. We say that the MAN scheme attains a {\it coded caching gain} of $t+1$ for $t=\Ksf\Msf/\Nsf$.
%In~\cite{ontheoptimality}, it was showed that MAN scheme is optimal under the constraint of uncoded cache placement when $\Ksf \leq \Nsf$. 
%In~\cite{yas2}, a variation of the MAN scheme (optimal under the constraint of uncoded cache placement when $\Ksf > \Nsf$) was shown to be information theoretically optimal to within a factor $2$ for shared-link networks.
A slight variation of the MAN scheme is known to be at most a factor of $2$ from an information theoretical outer bound~\cite{yas2}.

\begin{figure}%[ht]
%\vspace{-2mm}
\centerline{\includegraphics[scale=0.16]{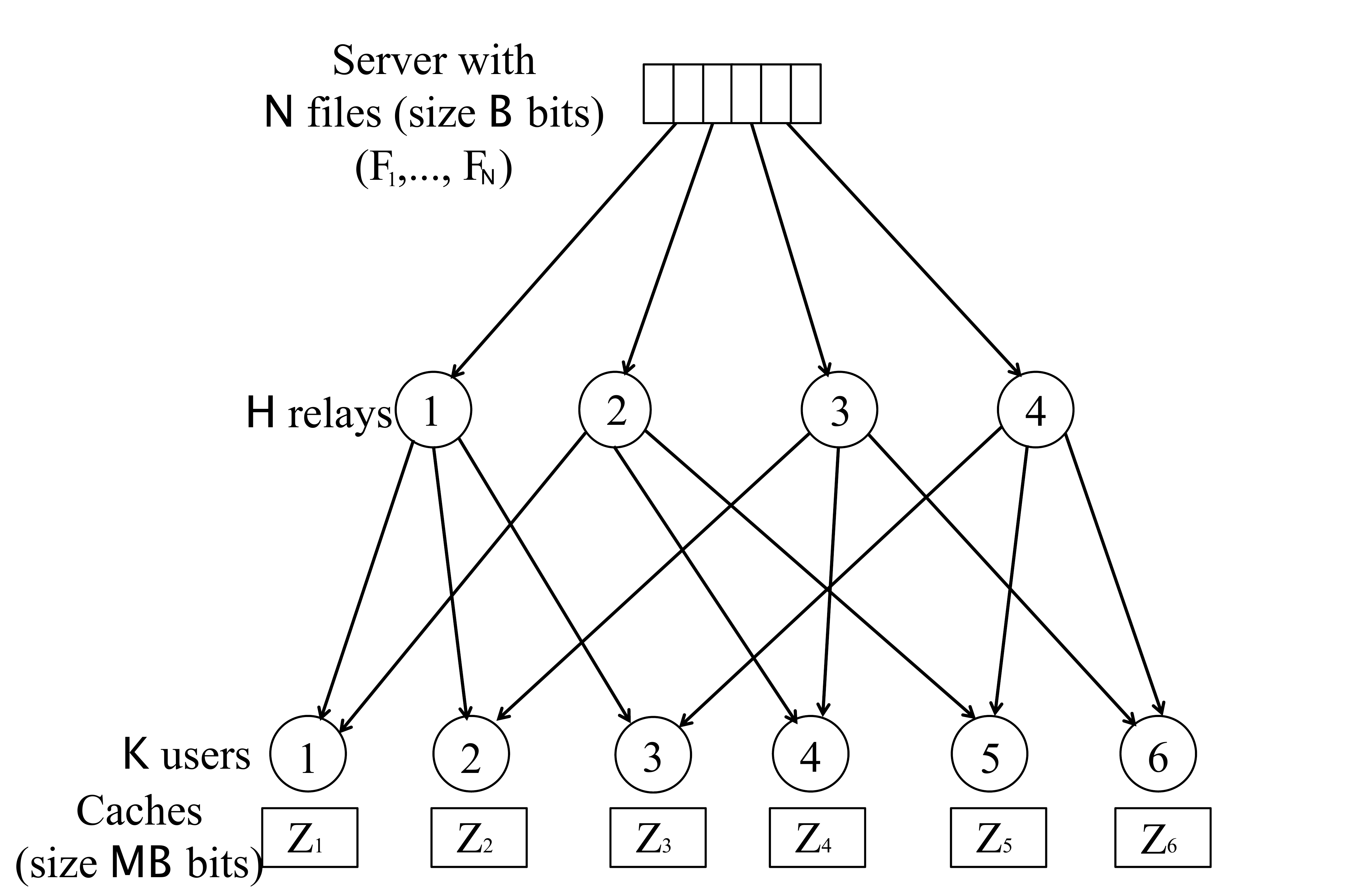}}
\caption{\small A combination network with $\Hsf=4$ relays and $\Ksf=6$ users, i.e., $\rsf=2$.}
\label{fig: Combination_Networks}
\vspace{-5mm}
\end{figure}

\paragraph*{\textbf{Combination networks}}
In practice, users may communicate with the central server through intermediate relays.
%in~\cite{multiserver}.
Since it is difficult to analyze general relay networks, a symmetric network, known as {\it combination network}~\cite{cachingincom}, has received a significant attention recently. 
A $(\Hsf,\rsf,\Msf,\Nsf)$ combination network comprises a server with $\Nsf$ files that is connected to $\Hsf$ relays (without caches) through $\Hsf$ orthogonal links, and each of the $\Ksf := \binom{\Hsf}{\rsf}$ users (with caches of size $\Msf$ files) is connected to a different subset of  $\rsf$ relays through $\rsf$ orthogonal links%(each relay is connected to $\binom{\Hsf-1}{\rsf-1}$ users)
--see Fig.~\ref{fig: Combination_Networks}.  The goal is to design a two-phase caching scheme that attains the {\it max-link-load}, that is, that minimizes the maximum number of transmissions among all links, which is related to the download time.

Past work can be divided into two groups.
%, depending on whether {\it uncoded} or {\it coded} cache placement is used in the achievable scheme.

\paragraph*{\textbf{Past work for combination networks with uncoded placement}}
With MAN placement and MAN multicast message generation, the authors in~\cite{cachingincom,novelwan2017} proposed various delivery schemes.
%
%Different to the previous  {MAN} placement based schemes, 
The scheme in~\cite{wan2017novelmulticase} still used MAN placement but proposed a novel way to generate and to deliver the multicast messages by leveraging the symmetries in the network topology.
Placement Delivery Array (PDA), originally proposed in~\cite{ontheplacementarrray} to reduce the sub-packetization of the {MAN} scheme in the shared-link model, has been recently extended in~\cite{PDA2017yan} to combination network; when $\rsf$ divides $\Hsf$, the scheme achieves the same load as%~\cite{cachingincom} and
~\cite{Zewail2017codedcaching} but with lower sub-packetization and with uncoded placement. 

The main limitation of schemes based on MAN placement is that, due to the combination network topology, 
the ``multicast opportunities'' (directly related to the overall coded caching gain) to transmit the various subfiles are different across subfiles. %, e.g., it may require more bits to transmit some subfiles than the other ones. 
Hence, even if the placement is symmetric, the %transmissions of subfiles are asymmetric.
delivery may be asymmetric. Since worst-case performance is of interest here, asymmetric delivery schemes are not desirable and they may actually be suboptimal.

\paragraph*{\textbf{Past work for combination networks with coded placement}}
In~\cite{asymmetric2018wan} we showed that coded placement schemes can be strictly better than any possible scheme with uncoded placement. 
The authors in~\cite{Zewail2017codedcaching} proposed a caching scheme where an MDS code is used before (symmetric) placement  so that the delivery phase for the combination network is equivalent to the delivery phase of $\Hsf$ uncoordinated shared-link networks, each serving $\binom{\Hsf-1}{\rsf-1}$ virtual users.
%The limitation of this coded placement scheme is that the coded caching gain is now that of a network with $\binom{\Hsf-1}{\rsf-1} < \Ksf=\binom{\Hsf}{\rsf}$ equivalent users, which appears to be suboptimal in light of known results for shared-link networks (i.e., the coded caching gain {\it fundamentally} scales linearly with the number of users $\Ksf$).

Our recent results in~\cite{asymmetric2018wan} used {\it asymmetric coded placement} with an MDS precoding to further reduce the max-link load achieved by~\cite{Zewail2017codedcaching} when the cache size is large; the MDS code parameters are not the same in the two papers.
The key idea in~\cite{asymmetric2018wan} is to let the users decode only those subfiles that can be transmitted with other $g-1$ equal-length subfiles in a single linear combination from a single relay;
the main drawback is that when $g$ (i.e., and thus the cache size) is small some multicasting opportunities are ``overlooked.''

\paragraph*{\textbf{Contributions}}
In this paper %, as opposed to~\cite{Zewail2017codedcaching}, 
we design an asymmetric coded placement so that the delivery by the $\Hsf$ relays can be ``coordinated''--to be made precise later.
We also prove that the proposed schemes strictly lower the max-link-load compared to~\cite{Zewail2017codedcaching} when $g\leq \binom{\Hsf-2}{\rsf-2}+1$.
Numerical evaluations show that the proposed schemes outperform existing schemes.

\paragraph*{\textbf{Paper Organization}}
The paper is organized as follows.
Section~\ref{sec:model} gives the formal problem definition and some related results.
Section~\ref{sec:baseline scheme} states the main results. 
%Section~\ref{sec:numerical results} presents numerical evaluations and
Section~\ref{sec:conclusion} concludes this paper.

\section{System Model and Related Results}
\label{sec:model}

\subsection{Notation}
\label{sub:notation}
We shall use the following notation convention in the study of the $(\Hsf,\rsf,\Msf,\Nsf)$ combination network, where a server with $\Nsf$ files communicates with the users through $\Hsf$ immediate relays, and each user has local cache of size $\Msf$ files and is connected to a different subset of $\rsf$ relays. 
We let 
\begin{align}
\Ksf_{i}:=\binom{\Hsf-i}{\rsf-i}, \ i\in [0:\rsf],
\label{eq:def of Ki}
\end{align}
where 
$\Ksf_0=\Ksf$ is the number of users in the system,
$\Ksf_1$ is the number of users connected to each relay,
and $\Ksf_i$ represents the number of users that are simultaneously connected to $i$ relays.
Our convention is that $\binom{x}{y}=0$ if $x<0$ or $y<0$ or $x<y$.

The subset of users connected to relay $h\in[\Hsf]$ is denoted by $\Uc_{h}$, and
the subset of relays connected to user $k\in[\Ksf]$ %is denoted
by $\Hc_{k}$.
For a subset of users $\Wc\subseteq[\Ksf]$, the set of relays simultaneoulsy connected to all the users in $\Wc$ is denoted by 
\begin{align}
\Rc_{\Wc}:=\{h\in[\Hsf]: \Wc \subseteq \Uc_h \}.
\label{eq:RW def}
\end{align}
For a subset of relays $\Jc\subseteq [\Hsf]$, the set of users who are simultaneously connected to all the relays in $\Yc$ is denoted by 
\begin{align}
 \Uc_{\Yc}:=\{k\in [\Ksf]: k\in\cap_{h\in\Jc} \ \Uc_h\}.
\label{eq:PY def}
\end{align}
Note that $\Uc_{\{h\}} = \Uc_h$.
For a given integer $t$, the $t$-subsets of users for which there exists at least one relay connected to all the users in this subset is denoted as
\begin{align}
\Zc_t:=\big\{\Wc\subseteq [\Ksf]:|\Wc|=t, \ \Rc_{\Wc}\neq\emptyset\big\}.
\label{eq:def of Z}
\end{align}
By the inclusion-exclusion principle~\cite[Theorem~10.1]{combinatorics}
\begin{align}
|\Zc_{t}|=\sum_{n=1}^{\rsf}\binom{\Hsf}{n}\binom{\Ksf_n}{t}(-1)^{n-1},
\label{eq:number of Zt}
\end{align}
and moreover, from the definition of $\Ksf_i$ in~\eqref{eq:def of Ki}, we have 
\begin{align}
\frac{t|\Zc_{t}|}{\Ksf_0}
  &=\sum_{n=1}^{\rsf} \binom{\Hsf}{n} \frac{\Ksf_n}{\Ksf_0} \binom{\Ksf_n-1}{t-1} (-1)^{n-1}
\notag
\\&=\sum_{n=1}^{\rsf} \binom{\rsf}{n}  \binom{\Ksf_n-1}{t-1} (-1)^{n-1}.
\label{eq:number of t*Zt}
\end{align}
For the network in~Fig.~\ref{fig: Combination_Networks}, we have
\begin{align*}
 &\Uc_{1}=\{1,2,3\}, \ 
  \Uc_{2}=\{1,4,5\}, 
  \\ 
 &\Uc_{3}=\{2,4,6\}, \ 
  \Uc_{4}=\{3,5,6\}.
\end{align*}
and thus, for instance, 
%$\Uc_{1}=\{1,2,3\}$,
%$\Hc_{1}=\{1,2\}$, 
$\Rc_{\{1,2\}}=\{1\}$, 
$\Uc_{\{2,3\}}=\Uc_2\cap \Uc_3 = \{4\}$, and
$\Zc_{1}$ contains all the $1$-subsets of $[6]$, while
$\Zc_{2}$ contains all the $2$-subsets of $[6]$ with the exception of $\{1,6\},\{2,5\},\{3,4\}$.

Moreover, 
calligraphic symbols denote sets or collections (i.e., set of sets), 
bold symbols denote vectors, 
and sans-serif symbols denote system parameters.
We use $|\cdot|$ to represent the cardinality of a set or the absolute value of a real number;
$[a:b]:=\left\{ a,a+1,\ldots,b\right\}$ and $[n] := [1:n]$; 
$\oplus$ represents bit-wise XOR. 
%We define the set
%\begin{align}
%\arg \max_{x\in \Xc}f(x) := \big\{x\in \Xc:f(x)=\max_{x^{\prime}\in\Xc}f(x^{\prime})\big\}.
%\label{eq:def of argmaxfunct}
%\end{align}

\subsection{System Model}
\label{sub:system model}
In a $(\Hsf,\rsf,\Msf,\Nsf)$ combination network, a server has $\Nsf$ files, denoted by $F_1, \cdots, F_\Nsf$, each composed of $\Bsf$ i.i.d uniformly distributed bits.
The server is connected to $\Hsf$ relays through $\Hsf$ error-free orthogonal links. 
The relays are connected to $\Ksf := \Ksf_0$ users through $\rsf \, \Ksf$ error-free orthogonal links.
Each user has a local cache of size $\Msf\Bsf$ bits, for $\Msf\in[0,\Nsf]$,
and is connected to a distinct $\rsf$-subset of relays.

% placement
In the placement phase, user $k\in[\Ksf]$ stores information about the $\Nsf$ files in its cache of size $\mathsf{MB}$ bits, where $\Msf \in[0,\Nsf]$.  %This phase is done without knowledge of users' demands. 
The cache content of user $k\in[\Ksf]$ is denoted by $Z_{k}$; let $\Zm:=(Z_{1},\ldots,Z_{\Ksf})$.
% demands
During the delivery phase, user $k\in[\Ksf]$ requests file $d_{k}\in[\Nsf]$;
the demand vector $\dv:=(d_{1},\ldots,d_{\Ksf})$ is revealed to all nodes. 
% delivery
Given $(\dv,\Zm)$, the server sends a message $X_{h}$ 
of $\Bsf \, \Rsf_{h}(\dv,\Zm)$ bits to relay $h\in [\Hsf]$. 
Then, relay $h\in [\Hsf]$ transmits a message $X_{h\to k}$ 
of $\Bsf \, \Rsf_{h\to k}(\dv,\Zm)$ bits to user $k \in \Uc_h$. 
% decoding
User $k\in[\Ksf]$ must recover its desired file $F_{d_{k}}$ from $Z_{k}$ and $(X_{h\to k} : h\in \Hc_k)$ with high probability when $\Bsf\to \infty$. 
% goal
The {\it max-link load} $\Rsf^{\star}$ is 
\begin{align}%{rCl}
\Rsf^{\star} &:=
\min_{\substack{\Zm}} 
\max_{\dv\in[\Nsf]^{\Ksf}} 
\left\{
\Rsf_1(\dv,\Zm), 
\Rsf_2(\dv,\Zm)
\right\},
\label{eq:def Rstar}
\\
\Rsf_1(\dv,\Zm) &:= \max_{h\in[\Hsf]} \{\Rsf_h(\dv,\Zm)\},
\label{eq:def R1}
\\
\Rsf_2(\dv,\Zm) &:= \max_{k\in\Uc_h, h\in[\Hsf]} \{\Rsf_{h\to k}(\dv,\Zm)\},
\label{eq:def R2}
\end{align}
where $\Rsf_1$ in~\eqref{eq:def R1} is the largest load from the server to the relays,
and   $\Rsf_2$ in~\eqref{eq:def R2} is the largest load from the relays to the users.

We say that a scheme with max-link load $\Rsf$ attains a {\it coded caching gain} of $g$ if
\begin{align}
%g:=\frac{\Rsf_{\rsf}}{\Rsf}=\frac{\Ksf/\Hsf (1-\Msf/\Nsf)}{\Rsf}. 
\Rsf
& = \frac{\Rsf_\text{routing}}{g}, \ \text{for} \ 
\label{eq:def g}
\\
\Rsf_\text{routing} 
&:= \frac{\Ksf(1-\Msf/\Nsf)}{\Hsf}
  = \frac{\Ksf_1(1-\Msf/\Nsf)}{\rsf} \ \text{from~\cite{cachingincom}.}
\label{eq:def Rrouting}
\end{align}
%Recall that . 
By the cut-set bound~\cite{cachingincom} we have $g\leq \Ksf_1 = \rsf\Ksf/\Hsf$ (recall that $\Ksf_1$ is the number of users connected to each relay).

\subsection{Caching Scheme in~\cite[Theorem~1]{Zewail2017codedcaching}}
\label{sub:caching scheme of Yener}
We state here the state-of-the-art scheme in~\cite{Zewail2017codedcaching} for the case of no cache at the relays;
the scheme uses MDS-based coded placement so as the delivery from each relay is equivalent to that of  
%treats the combination network into $\Hsf$ (uncoordinated) 
a shared-link network serving $\Ksf_1$ virtual users and where the operations of the $\Hsf$ virtual shared-link network are not coordinated.
In particular, each file %$F_i$ of $\Bsf$~bits, $i\in[\Nsf]$, 
is divided into $\rsf$ non-overlapping and equal-length pieces %of size $\Bsf/\rsf$ bits, which 
that are encoded by an $(\Hsf,\rsf)$ MDS code. 
The $h$-th MDS-coded symbol is denoted by $s^{h}_{i}$ and must be delivered 
%of size $|s^{h}_{i}|=\Bsf/\rsf$ for $h\in[\Hsf]$; 
by relay $h\in[\Hsf]$ to the users in $\Uc_h$ following the MAN scheme~\cite{dvbt2fundamental}. 
This is done as follows.

\paragraph*{Placement}
Fix $g\in [1:\Ksf_1]$.
The MDS-coded symbol $s^{h}_{i}$ is partitioned into $\binom{\Ksf_1}{g-1}$ non-overlapping and equal-length {\it subfiles} as $s^{h}_{i}=\{s^{h}_{i,\Wc}:\Wc\subseteq \Uc_h, \ |\Wc|=g-1\}$
%where each subfile has $\frac{\Bsf}{\rsf \binom{\Ksf_1}{g-1}}$~bits 
(recall $|\Uc_h|=\Ksf_1$ for all $h\in[\Hsf]$). 
There are in total
\begin{align}
\nsf = \Hsf\binom{\Ksf_1}{g-1} \ \text{[subfiles per file]}.
\label{eq:Yener n}
\end{align}
User $k\in[\Ksf]$ caches $s^{h}_{i,\Wc}$ if $k\in\Wc$ from all $h\in \Hc_{k}$
(recall $|\Hc_{k}|=\rsf$ for all users), for a total of
\begin{align}
\ksf_1 = \rsf \binom{\Ksf_1-1}{g-2} \ \text{[subfiles per file]}.
\label{eq:Yener k1}
\end{align}

\paragraph*{Delivery}
The MAN-like multicast coded message
\begin{align}
w^{h}_{\Jc}=\underset{k\in\Jc}{\oplus}s^{h}_{ d_k,\Jc\setminus \{k\}}, 
\ \forall \Jc\subseteq \Uc_h : |\Jc|=g, \ h\in [\Hsf],
\label{eq:Yener multicast coded messages}
\end{align}
is delivered from the server to relay $h$, who then forwards it to the users in $\Jc$. 
User $k\in[\Ksf]$, thanks to its cache content and the received multicast coded messages from the relays in $\Hc_{k}$, recovers
\begin{align}
\ksf_2 = \rsf \binom{\Ksf_1-1}{g-1} \ \text{[subfiles per file]}.
\label{eq:Yener k2}
\end{align}
Note that there are
\begin{align}
\ksf_3 = \Hsf\binom{\Ksf_1}{g} \ \text{[subfiles]}, 
\label{eq:Yener k3}
\end{align}
multicast coded messages in~\eqref{eq:Yener multicast coded messages}, each of the size of a subfile, that are delivered from the server to the relays.

\paragraph*{Performance}
Each user eventually knows $\ksf_1+\ksf_2 = \rsf \binom{\Ksf_1}{g-1}$ subfiles of its desired file (either cached or delivered), which suffices to recover all the $\nsf = \Hsf\binom{\Ksf_1}{g-1}$ subfiles of its desired file  because of the $(\Hsf,\rsf)$ MDS encoding before placement, where $\ksf_1$, $\ksf_2$ and $\nsf$ are defined in~\eqref{eq:Yener k1},~\eqref{eq:Yener k2} and~\eqref{eq:Yener n}, respectively.
Since each multicast coded message in~\eqref{eq:Yener multicast coded messages} is simultaneously useful for $g$ users, a coded caching gain of $g$ is achieved and the required memory size is
\begin{align}
\Msf 
 = \Nsf \frac{\ksf_1}{\nsf} \cdot \frac{\Hsf}{\rsf}
%:=\Nsf\rsf  \frac{\binom{\Ksf_1-1}{g-2}}{\rsf\binom{\Ksf_1}{g-1}}
 =\Nsf \frac{g-1}{\Ksf_1}
 =: \Msf_\text{\rm\cite{Zewail2017codedcaching}}(g),
\label{eq:Yener memory}
\end{align}
where in~\eqref{eq:Yener memory} the factor $\frac{\Hsf}{\rsf}$ is the inverse of the rate of the MDS code used before placement.

In general, the used MDS code has parameters $(\nsf,\ksf_1+\ksf_2)$ because each users must be able to recover $\nsf$ subfiles from the available $\ksf_1+\ksf_2$ subfiles;
therefore for a scheme where the delivery is symmetric across users and relays we have
\begin{align}
%\frac{1}{\eta} \cdot\frac{\ksf_1}{\nsf} 
\frac{\Msf}{\Nsf} &= \frac{\ksf_1}{\ksf_1+\ksf_2}
\ \text{(memory occupancy per file)},
\label{eq:k1/n def}
\\
%\frac{1}{\eta} \cdot\frac{\ksf_2}{\nsf} 
\rsf\Rsf_2 
&= \frac{\ksf_2}{\ksf_1+\ksf_2}
= \left(1-\frac{\Msf}{\Nsf}\right)
\ \text{(total load to a user)},
\label{eq:k2/n def}
\\
%\frac{1}{\eta} \cdot\frac{\ksf_3}{\nsf} 
\Hsf\Rsf_1
&= \frac{\ksf_3}{\ksf_1+\ksf_2}
 = \frac{\Hsf \ksf_3}{\Ksf \ksf_2} \Rsf_\text{routing}
\ \text{(load to the relays)},
\label{eq:k3/n def}
\\
\Longleftrightarrow 
g &= \frac{\Ksf \ksf_2}{\ksf_3} %\  \Rsf=\max(\Rsf_1,\Rsf_2)=\Rsf_1,
\ \text{(coded caching gain)},
\end{align}
%in~\eqref{eq:k2/n def} because the $\ksf_2$ subfiles per file are delivered across  $\rsf$ links to each user, and
%in~\eqref{eq:k3/n def} because the $\ksf_3$ multicast coded messages are delivered across  $\Hsf$ links.
where $\Rsf_1$ and $\Rsf_2$ were defined in~\eqref{eq:def R1} and~\eqref{eq:def R2}, respectively;
notice that $\Ksf \ksf_2$ represents the total number of subfiles decoded by the users and $\ksf_3$ is the number of subfiles actually sent.

\paragraph*{Limitation}
In~\cite{Zewail2017codedcaching}, %the combination network is treated as 
the operations at the $\Hsf$ relays are uncoordinated.
%\begin{example}[$\Hsf=4$, $\rsf=2$, $\Nsf=6$, $g=2$]
%\label{ex:yener}
Indeed, consider the network in Fig.~\ref{fig: Combination_Networks}  for $g=2$.
The scheme in~\cite{Zewail2017codedcaching} %treats the combination network as $\Hsf$ uncoordinated shared-link models by 
uses an $(\Hsf,\rsf)=(4,2)$ MDS code, and the MDS-coded symbols $s^{h_1}_{i,\Wc}$ and $s^{h_2}_{i,\Wc}$ are treated as two ``independent'' subfiles if $h_1\neq h_2$. 
For example, among the MDS subfiles 
$s^{1}_{i,\{1\}}$, $s^{1}_{i,\{2\}}$, $s^{1}_{i,\{3\}}$, 
$s^{2}_{i,\{1\}}$, $s^{2}_{i,\{4\}}$ and $s^{2}_{i,\{5\}}$,
each of length is $\Bsf/6$, 
user~1 caches $s^{1}_{i,\{1\}}$ and $s^{2}_{i,\{1\}}$, %with totally $2\Bsf/6$. So the needed memory size is 
which requires $\Msf/\Nsf=2/6$. %_\text{\rm(\cite{Zewail2017codedcaching})}(g)
However, $s^{1}_{i,\{1\}}$ and $s^{2}_{i,\{1\}}$ can be treated as a single subfile known / cached by user~1.
%, denoted by $f_{i,\{1\}}$; by using an $(6,5)$ MDS code, among the ``independent'' MDS-coded symbols $f_{i,\{1\}}$,  $f_{i,\{2\}}$,  $f_{i,\{3\}}$,  $f_{i,\{4\}}$ and  $f_{i,\{5\}}$, each of length is $\Bsf/5$, user~1 caches $f_{i,\{1\}}$ so the needed memory size is only $\Msf/\Nsf=1/5$.
%
This observation is key for the design of the novel proposed schemes. %described in Section~\ref{sec:proof of thm 1}.
%\end{example}

\section{Main Result}
\label{sec:baseline scheme}
In this section, we describe the proposed scheme that aims to overcome the limitation of~\cite[Theorem~1]{Zewail2017codedcaching} as discussed in the previous section. We have:
\begin{thm}
\label{thm:baseline scheme}
For an $(\Hsf,\rsf,\Msf,\Nsf)$ combination network, 
%the lower convex envelop of the following points
a coded caching gain $g\in [1:\Ksf_1]$ is achievable
with a memory requirement of
\begin{align}
\Msf 
 =
\Nsf\frac{\sum_{a=1}^{\rsf}\binom{\rsf}{a}\binom{\Ksf_a-1}{g-2}(-1)^{a-1}}{\sum_{a=1}^{\rsf}\binom{\rsf}{a}\binom{\Ksf_a}{g-1}(-1)^{a-1}}
=: \Msf_\text{\rm[Th.1]}(g).
\label{eq:our baseline scheme}
\end{align}
%The tradeoff between memory size and max-link load is the lower convex envelope of the above points.
\end{thm}

%We use the rest of the section to prove Theorem~\ref{thm:baseline scheme}.
%
%\subsection{Proof of Theorem~\ref{thm:baseline scheme}}
%\label{sec:proof of thm 1}

\begin{IEEEproof}
We aim to achieve coded caching gain $g$. In other words, every multicast coded message send through the network is simultaneously useful for $g$ users and each subfile is cached by at least $g-1$ other users.
 
\paragraph*{Placement}
We consider the elements of $\Zc_{g-1}$ defined in~\eqref{eq:def of Z}, that is, those subsets of users with cardinality $g-1$ (from a ground set of cardinality $\Ksf_1$) for which there exists at least one relay connected to all of them. 
%for $\Rc_{\Wc}$ defined in~\eqref{eq:RW def}. 
We aim to partition each {\it MDS-coded file} into 
\begin{align}
\nsf=|\Zc_{g-1}| \ \text{[subfiles per file]}\label{eq:us1 n}
\end{align}
equal-length subfiles, i.e., 
$f_i = (f_{i,\Wc} : \Wc\in\Zc_{g-1}),  \ i\in[\Nsf],$ where subfile $f_{i,\Wc}$ 
%has size $|f_{i,\Wc}| = \ssf$~bits and 
is cached by the users in $\Wc$.
Therefore, each user caches 
\begin{align}
\ksf_1 = \frac{g-1}{\Ksf} |\Zc_{g-1}| \ \text{[subfiles per file]},
\label{eq:us1 k1}
\end{align}
since each subfile is cached by $g-1$ users and all users cache the same amount of subfiles.
This placement is considered to be {\it asymmetric} because not all subfiles $f_{i,\Wc}$ for $\Wc \subseteq[\Ksf]$ of cardinality $|\Wc|=g-1$ are present.

\paragraph*{Delivery}
We should create  a multicast coded message similarly to~\eqref{eq:Yener multicast coded messages}
for each subset of users $\Jc$ of the form 
\begin{align}
\Jc = \Wc \cup \{k\} :  \Wc\in \Zc_{g-1}, \ k\in[\Ksf], \ k \not\in\Wc;
\label{eq:usJmust}
\end{align} 
%there are $\Ksf|\Zc_{g-1}|$ different sets in~\eqref{eq:usJmust}.
%
however, only those $\Jc\in\Zc_{g}$ %, for $\Jc$ as in~\eqref{eq:usJmust}, 
are such that all users in $\Jc$ have at least one common connected relay; in order to have a symmetric delivery scheme from the relays to the users, we aim to deliver only those multicast coded messages for $\Jc\in \Zc_{g}$ and consider those for $\Jc\not\in \Zc_{g}$ as ``erased'', i.e., $\ksf_3 = |\Zc_{g}|$. Therefore, each user eventually decodes
\begin{align}
\ksf_2=\frac{g|\Zc_{g}|}{\Ksf} \ \text{[subfiles per file]},
\label{eq:us1 k2}
\end{align}
More precisely, for each set $\Jc\in\Zc_g$, we generate the MAN-like multicast message 
\begin{align}
W_{\Jc} =\underset{k\in\Jc}{\oplus}f_{d_k,\Jc\setminus \{k\}}.
\end{align} 
We then divide $W_{\Jc}$ into $|\Rc_{\Jc}|$ non-overlapping and equal-length pieces $W_{\Jc}=\{W^h_{\Jc}:h\in \Rc_{\Jc}\}$; the server transmits $W^h_{\Jc}$ to relay $h\in \Rc_{\Jc}$, which then forwards it to users in $\Jc$.
 A user must be able to recover all the $\nsf$ subfiles of its desired file 
from the $\ksf_1+\ksf_2$ subfiles that were either cached or received;
this is possible if we divide each file into $\ksf_1+\ksf_2$ non-overlapping and equal-length pieces and use 
an $(\nsf,\ksf_1+\ksf_2)$ MDS code to generate the subfiles before placement, where $\ksf_1$, $\ksf_2$ and $\nsf$ are defined in~\eqref{eq:us1 k1},~\eqref{eq:us1 k2} and~\eqref{eq:us1 n}, respectively.

\paragraph*{Performance}
By the above construction, each multicast coded message is simultaneously useful for $g$ users, thus a coded caching gain of $g$ is achieved with cache size (see~\eqref{eq:k1/n def})
\begin{align}
\frac{\Msf_\text{\rm[Th.1]}(g)}{\Nsf}
 &=\frac{\ksf_1}{\ksf_1+\ksf_2}
  =\frac{(g-1)|\Zc_{g-1}|}{(g-1)|\Zc_{g-1}|+g|\Zc_{g}|},
% =: \Msf_\text{\rm[Th.1]}(g),
\label{eq:us1 memory}
\end{align}
%The subscript `b' in $\Msf_\text{\rm[Th.1]}(g)$ stands for `baseline'.
By using~\eqref{eq:number of t*Zt} in~\eqref{eq:us1 memory}, and the identity 
$\binom{\Ksf_a-1}{g-2} + \binom{\Ksf_a-1}{g-1} = \binom{\Ksf_a}{g-1}$,
we obtained the claimed cache size in~\eqref{eq:our baseline scheme}.
\end{IEEEproof}

\subsection{Comparison between Theorem~\ref{thm:baseline scheme} and~\cite[Theorem~1]{Zewail2017codedcaching}}
In the following we show that our scheme in Theorem~\ref{thm:baseline scheme} is no worse than the scheme in~\cite{Zewail2017codedcaching}. 
%%%For example, it can be easily seen that for the combination network with $\Hsf=4$, $\rsf=2$, $\Nsf=\Ksf=6$, {\red and $\Msf=6/5$ WE SHOULD FIX g!!!} in Fig.~\ref{fig: Combination_Networks},
%%%%Each demanded MDS subfile  is transmit in one linear combination which also includes other $g-1=1$ demanded MDS subfiles with identical length and thus the coded caching gain is $g=2$.
%%%the achieved max-link load with Theorem~\ref{thm:baseline scheme} is $3/5$, which coincides with the cut-set outer bound based on~\cite{yas2}, while the max-link load with the scheme from~\cite{Zewail2017codedcaching} is $9/10$.
In general we have:

\begin{cor}
\label{cor:comparison Th1 to Yener} 
For an $(\Hsf,\rsf,\Msf,\Nsf)$ combination network with coded caching gain $g\in[\Ksf_1]$, %it holds that 
$\Msf_\text{\rm[Th.1]}(g) \leq \Msf_\text{\rm\cite{Zewail2017codedcaching}}(g)$ with equality if and only if $g \geq \Ksf_2+2$.
\end{cor}
\begin{IEEEproof}
The proof %is really simple and 
uses the fact that $\Ksf_{\rsf} < \Ksf_{\rsf-1} \ldots < \Ksf_{1}$.
Indeed, $\Msf_\text{\rm[Th.1]}(g)$ in~\eqref{eq:us1 memory} %{eq:our baseline scheme}
is no larger than $\Msf_\text{\rm\cite{Zewail2017codedcaching}}(g)$ in~\eqref{eq:Yener memory} if
$
\frac{|\Zc_{g}|}{|\Zc_{g-1}|}
\leq \frac{\Ksf_{1}-g+1}{g},
$
which is always true because
\begin{align}
\frac{|\Zc_{g}|}{|\Zc_{g-1}|}
  &=\frac
{\sum_{n=1}^{\rsf}\binom{\Hsf}{n}\binom{\Ksf_n}{g}(-1)^{n-1}}
{\sum_{n=1}^{\rsf}\binom{\Hsf}{n}\binom{\Ksf_n}{g-1}(-1)^{n-1}}
\label{eq:Th1 vs ZT: step2}
\\&= 
\frac
{\sum_{n=1}^{\rsf}\binom{\Hsf}{n}\frac{\Ksf_n-g+1}{g}\binom{\Ksf_n}{g-1}(-1)^{n-1}}
{\sum_{n=1}^{\rsf}\binom{\Hsf}{n}\binom{\Ksf_n}{g-1}(-1)^{n-1}}
\label{eq:Th1 vs ZT: step3}
\\&\leq
\frac
{\sum_{n=1}^{\rsf}\binom{\Hsf}{n}\frac{\Ksf_1-g+1}{g}\binom{\Ksf_n}{g-1}(-1)^{n-1}}
{\sum_{n=1}^{\rsf}\binom{\Hsf}{n}\binom{\Ksf_n}{g-1}(-1)^{n-1}}
\label{eq:Th1 vs ZT: step4}
\\&= \frac{\Ksf_{1}-g+1}{g}.
\label{eq:Th1 vs ZT: step5}
\end{align}
Moreover equality holds in~\eqref{eq:Th1 vs ZT: step4} if and only if the summations contain only one term, which is the case if and only if $\Ksf_2 < g-1$ (i.e., $\binom{\Ksf_2}{g-1}=0$) as claimed.
\end{IEEEproof}

\subsection{Numerical Results}
\label{sec:numerical results}

\begin{figure}%[tbh]
\centering{}
\includegraphics[scale=0.65]{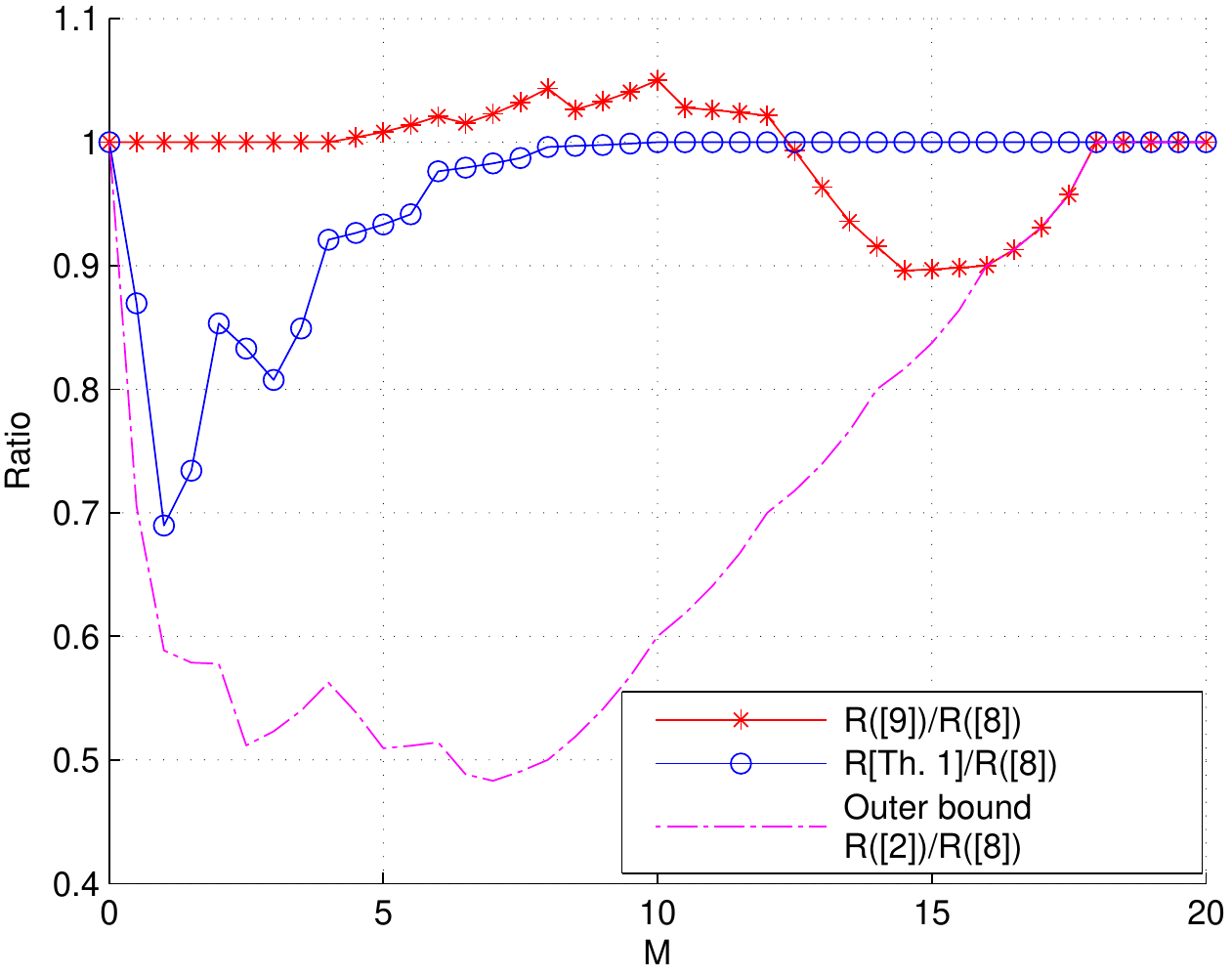}
\caption{Performance comparison for the combination network with $\Hsf=6$, $\rsf=3$ and $\Ksf=\Nsf=20$.}
\label{fig:numerical 1}
\end{figure}

In Fig.~\ref{fig:numerical 1}, we compare the performance of the proposed schemes to those of the schemes with coded cache placement in~\cite{Zewail2017codedcaching} and~\cite{asymmetric2018wan}. 
As an outer bound, we use the same cut-set idea of~\cite{cachingincom} (which used the cut-set bound for the shared-link model originally proposed in~\cite{dvbt2fundamental}) but with the enhanced cut-set for the shared-link model in~\cite{yas2}; we denote this outer bound as $\Rsf^\text{\rm(out)}_\text{\rm\cite{yas2}}$. 
In Fig.~\ref{fig:numerical 1}, we plot the ratios 
$\Rsf_\text{\rm\cite{asymmetric2018wan}}/\Rsf_\text{\rm\cite{Zewail2017codedcaching}}$ (red line), 
$\Rsf_\text{\rm[Th.1]}/\Rsf_\text{\rm\cite{Zewail2017codedcaching}}$ (blue line), and 
$\Rsf^\text{\rm(out)}_\text{\rm\cite{yas2}}/\Rsf_\text{\rm\cite{Zewail2017codedcaching}}$ (magenta dotted line), 
and where 
$\Rsf_\text{\rm\cite{asymmetric2018wan}}$ and
$\Rsf_\text{\rm\cite{Zewail2017codedcaching}}$ are the achievable max-link load by the schemes in~\cite{asymmetric2018wan} and in~\cite{Zewail2017codedcaching}, respectively.
We plot the ratio of max-link loads as otherwise their difference would not be clearly visible on a small figure.
It can be noted from Fig.~\ref{fig:numerical 1} that the blue curve, which represents our proposed scheme in Theorem~\ref{thm:baseline scheme}, is never below one, that is, it is never inferior in performance to the baseline scheme in~\cite{Zewail2017codedcaching};
however, it is strictly worse than the performance of our past work in~\cite{asymmetric2018wan} for $\Msf>12.5$ (which is information theoretically optimal for $\Msf\geq 16$).  Our proposed scheme in Theorem~\ref{thm:baseline scheme} is information theoretically optimal for $\Msf\geq 18$ and has the same max-link load as the scheme in~\cite{Zewail2017codedcaching}.

%RED = large g
%BLUE = small g
%need g > K2
%no constant gap , seems to depend on H
From Fig.~\ref{fig:numerical 1} we observe a general fenomenon: 
our scheme in Theorem~\ref{thm:baseline scheme} (blue line) improves on the scheme in~\cite{Zewail2017codedcaching} for small value of $g$, while our scheme in~\cite{asymmetric2018wan} (red line) improves on the scheme in~\cite{Zewail2017codedcaching} for large value of $g$. Part of our ongoing work is to design a scheme that combines the advantages of both Theorem~\ref{thm:baseline scheme} and~\cite{asymmetric2018wan}. 
In Corollary~\ref{cor:comparison Th1 to Yener} we proved that  Theorem~\ref{thm:baseline scheme} is equivalent to the scheme in~\cite{Zewail2017codedcaching} for $g \geq \Ksf_2+2$; this suggests that an improved scheme should consider the multicasting coding opportunities for groups of $\Ksf_2$ users or more.

Finally, numerical evaluations suggest that the ratio $\Rsf^\text{\rm(out)}_\text{\rm\cite{yas2}}/\Rsf_\text{\rm[Th.1]}$ is increases as $\Hsf$ increases. An interesting open question is thus if any of the known achievable schemes is to within a constant factor of a known outer bound.

\section{Conclusions}
\label{sec:conclusion}
This paper proposed a novel asymmetric coded cache placement scheme for combination networks with end-user-caches, which aim to create multicasting opportunities across relays. The proposed schemes were shown to be achieve a max-link load no larger than the best scheme known in the literature.

\section*{Acknowledgment}
This work was supported in parts by NSF 1527059 and Labex DigiCosme.

\newpage
\bibliographystyle{IEEEtran}
\bibliography{IEEEabrv,IEEEexample}

\begin{thebibliography}{10}
\providecommand{\url}[1]{#1}
\csname url@samestyle\endcsname
\providecommand{\newblock}{\relax}
\providecommand{\bibinfo}[2]{#2}
\providecommand{\BIBentrySTDinterwordspacing}{\spaceskip=0pt\relax}
\providecommand{\BIBentryALTinterwordstretchfactor}{4}
\providecommand{\BIBentryALTinterwordspacing}{\spaceskip=\fontdimen2\font plus
\BIBentryALTinterwordstretchfactor\fontdimen3\font minus
  \fontdimen4\font\relax}
\providecommand{\BIBforeignlanguage}[2]{{%
\expandafter\ifx\csname l@#1\endcsname\relax
\typeout{** WARNING: IEEEtran.bst: No hyphenation pattern has been}%
\typeout{** loaded for the language `#1'. Using the pattern for}%
\typeout{** the default language instead.}%
\else
\language=\csname l@#1\endcsname
\fi
#2}}
\providecommand{\BIBdecl}{\relax}
\BIBdecl

\bibitem{dvbt2fundamental}
M.~A. Maddah-Ali and U.~Niesen, ``Fundamental limits of caching,'' \emph{IEEE
  Trans. Infor. Theory}, vol.~60, no.~5, pp. 2856--2867, May 2014.

\bibitem{yas2}
Q.~Yu, M.~A. Maddah-Ali, and S.~Avestimehr, ``Characterizing the rate-memory
  tradeoff in cache networks within a factor of 2,'' \emph{in IEEE Int. Symp.
  Inf. Theory}, Jun. 2017.

\bibitem{cachingincom}
M.~Ji, M.~F. Wong, A.~M. Tulino, J.~Llorca, G.~Caire, M.~Effros, and
  M.~Langberg, ``On the fundamental limits of caching in combination
  networks,'' \emph{IEEE 16th Int. Workshop on Sig. Processing Advances in
  Wireless Commun.}, pp. 695--699, 2015.

\bibitem{novelwan2017}
K.~Wan, M.~Ji, P.~Piantanida, and D.~Tuninetti, ``Novel outer bounds and inner
  bounds with uncoded cache placement for combination networks with
  end-user-caches,'' \emph{inner bounds in 55th Allerton Conf. Commun.,
  Control, Comp. , outer bounds in IEEE Inf. Theory Workshop 2017, available at
  arXiv:1701.06884v5}, Oct. 2017.

\bibitem{wan2017novelmulticase}
------, ``Caching in combination networks: Novel multicast message generation
  and delivery by leveraging the network topology,'' \emph{accepted to IEEE
  Intern. Conf. Commun (ICC 2018), available at arXiv:1710.06752}.

\bibitem{ontheplacementarrray}
Q.~Yan, M.~Cheng, X.~Tang, and Q.~Chen, ``On the placement delivery array
  design in centralized coded caching scheme,'' \emph{IEEE Trans. Infor.
  Theory}, vol.~63, no.~9, pp. 5821--5833, Sep. 2017.

\bibitem{PDA2017yan}
Q.~Yan, M.~Wigger, and S.~Yang, ``Placement delivery array design for
  combination networks with edge caching,'' \emph{arXiv:1801.03048}, Jan. 2018.

\bibitem{Zewail2017codedcaching}
A.~A. Zewail and A.~Yener, ``Coded caching for combination networks with
  cache-aided relays,'' \emph{in IEEE Int. Symp. Inf. Theory}, pp. 2438--2442,
  June 2017.

\bibitem{asymmetric2018wan}
K.~Wan, M.~Ji, P.~Piantanida, and D.~Tuninetti, ``On the benefits of asymmetric
  coded cache placement in combination networks with end-user caches,''
  \emph{submitted to IEEE Int. Symp. Inf. Theory}, Jan. 2018.

\bibitem{combinatorics}
J.~H.~V. Lint and R.~M. Wilson, ``A course in combinatorics (second edition),''
  \emph{Cambridge University Press, ISBN 9780521803403}, 2001.

\end{thebibliography}
\end{document}